# Simulating Genomes and Populations in the Mutation Space: An example with the evolution of HIV drug resistance.

Antonio Carvajal-Rodríguez

Departamento de Bioquímica, Genética e Inmunología. Universidad de Vigo, 36310 Vigo, Spain

E-mail address:
	AC-R: acraaj@uvigo.es

## Abstract

**Background**

When simulating biological populations under different evolutionary genetic models, backward or forward strategies can be followed. Backward simulations, also called coalescent-based simulations, are computationally very efficient. However, this framework imposes several limitations that forward simulation does not. In this work, a new simple and efficient model to perform forward simulation of populations and/or genomes is proposed. The basic idea considers an individual as the differences (mutations) between this individual and a reference or consensus genotype. Thus, this individual is no longer represented by its complete sequence or genotype. An example of the efficiency of the new model with respect to a more classical forward one is demonstrated. This example models the evolution of HIV resistance using the B_FR.HXB2 reference sequence to study the emergence of known resistance mutants to Zidovudine and Didanosine drugs.

**Results**

When representing individuals as mutations with respect to a wild genotype we obtain an improvement of several orders of magnitude in both computation space and time. This is due to the great amount of redundant information present in the genomes within populations. We depict the basic algorithms, mutation, recombination and fitness, needed to implement this kind of model. This sort of representation is appropriate to investigate properties of the viral quasispecieces theory. We demonstrate the model efficiency with an example of the evolution of drug resistance in HIV-1. The result obtained seems to corroborate that the evolution of resistance is extremely dependent on the population size and the progeny number of the virus. In addition this seems also to agree with the recently proposed idea that there is no universally applicable unique value of effective population size for HIV-1 but this will depend on the specific process of interest under study. In the case of the evolution of resistance in a short time period, it seems that no low or medium effective population sizes as $10^3 – 10^5$ should be assumed though in other situations it could.

**Conclusions**

We propose a forward simulation framework to represent individuals just as the mutations they carry with respect to the wild genotype. The new framework seems to be an efficient way for forward modeling genomes and/or populations in the computer. Taking advantage of the new modeling, we also show the importance of population sizes being large enough for HIV-1 to get resistance phenotypes faster in time.

# Background

There are many different situations in current population biology research where simulating populations at a genetic and/or ecological level is very useful. Some examples include exploring complex situations such as molecular clock-like evolution [1], the evolution of drug-resistance in HIV-1 [2], human impacts onto genetic diversity under different demographic scenarios [3], human populations undergoing complex diseases [4, 5], speciation processes [6], etc.

Simulation of populations also allows modeling spatially explicit situations as epidemiological ones [7] and different ecological and evolutionary scenarios with interest in conservation and management of populations [8, 9]. In addition, population simulation of genetic datasets allows getting expectations of parameters which are otherwise difficult to obtain such as genome-wide mutation rates [10] or the effect of deleterious mutational load on populations [11].

When simulating biological populations under different evolutionary genetic models, backward or forward strategies can be followed. Backward simulations, also called coalescent-based simulations, are computationally very efficient because they are backward based on the history of lineages with survived offspring in the current population ignoring, however, all those lineages whose offspring did not arrive to the present [12]. Hence, coalescence is a sample-based theory relevant to the study of population samples and DNA sequence data. Due to its efficiency, it has also been used to derive several algorithms to estimate parameter values that maximize the probability of the given data [13]. From its beginnings, the basic coalescence has been extended in several useful ways to include structured population models [14, 15] [16-18], changes in population size [19-21], recombination [22, 23] and selection [24-29].

By contrast, forward simulations are less efficient because the whole history of the sample is followed from past to present. However, the coalescent framework imposes some limitations not present in the forward simulation. The first of all is the same that causes its efficiency, namely, the coalescence does not keep track of the complete ancestral information. Thus, if the interest is focused on the evolutionary process itself, rather than on its outcome, forward simulations should be preferred [30]. Second, coalescent simulations are complicated by simple genetic forces such as selection, and although different evolutionary scenarios have been incorporated (see above) it is still difficult to implement models incorporating complex evolutionary situations with selection, variable population size, recombination, complex mating schemes, and so on. Similarly, coalescent methods cannot yet simulate realistic samples of complex human diseases [4]. Moreover, the coalescent model is based on specific limiting values and relationships between some important parameters [31]. Finally, no spatial explicit modelling is allowed within the coalescent framework.

In this work, I propose a new simple and efficient model to perform forward simulation of populations. The basic idea considers an individual as the differences (mutations) between this individual and a reference or consensus genotype. Thus, this individual is no longer represented by its complete genotype. An example of the efficiency of the new forward model with respect to a more classical forward one is demonstrated. This example models the evolution of HIV resistance using the B_FR.HXB2 reference sequence to study the appearance of known resistance mutants to Zidovudine and Didanosine drugs.

## Results

### Algorithm

Consider a forward simulation model of haploid organisms, e.g. proviruses, consisting of DNA or RNA sequences, with variable population size, constant selection and discrete generations. Let these organisms to reproduce inside host cells from which they are released as virions (diploids) that survive depending on their fitnesses values. These virions will enter the cells as proviruses (haploids) and will mutate and, if more than one provirus infected the cell, they will recombine to produce the next generation of released virions. This is a typical genetic population model of HIV life cycle [32-34].

Let the genome size to be of $10^4$ nucleotides and begin with one infected cell with one provirus. Then allow mutation with a rate μ per site per generation and release an $N$ number of virions that will survive according to their fitnesses. This process will be repeated infecting cells as proviruses, mutating, recombining and so on (see Methods for more details of the genetic population model).

In a classical forward simulator, the $N \times 10^4$ nucleotides corresponding to the sum of genome sizes in the population will be stored in the computer while they are evolving, mutating, surviving and reproducing during a number of generations. This is done at the cost of a lot of computation space and time. As the process starts with just one provirus, by storing all the genomes a great amount of redundant information will be stored. In contrast, it could be possible to keep the sequence information just once because all individuals except mutants will be identical. That is, if one mutation appears, this individual will be stored as a mutation position in the original sequence plus just the new mutant nucleotide or codon, depending on the genetic model being used, instead of the whole genome. Thus, the dimensionality of the problem will be reduced almost by a four–fold factor. If higher size genomes are considered with equal or lower mutation rate, the reduction of dimensionality will be increased. This way of representing individuals as mutant deviations from a reference sequence is called a representation in the mutation space (MS), compared to the classical one where individuals are represented by their full sequence (sequence space, SS). By using the MS representation, efficiency is gained in both computation space and time. However, there is also a necessity to redefine the implementation of some processes such as mutation, recombination and fitness evaluation to adjust to the new way of storing genomes in this less-redundant manner.

In what follows we will describe the implementation of the MS representation applied to the genetic population model above exposed. We will briefly describe the implementation of mutation, recombination, fitness evaluation and some possible extensions of the model. The C++ source code is available upon request from the author, and a new forward simulator using the MS representation will be soon freely available. Afterwards, we will perform an experiment as a test case, using a sequence DNA fragment (3012 bp) corresponding to the coding region of the subunit p51 of the reverse transcriptase from the consensus B_FR.HXB2 HIV-1 sequence, to study the evolution of resistance to Zidovudine and Didanosine drugs in a number *N* of virions through 250 generations under therapeutic conditions (see Methods section for details). The study will be performed by implementing the same model using two different forward simulators: 1) The conventional SS representation, and 2) The MS representation. The increase of fitness due to the emergence of resistant mutants and the mutation distribution will be evaluated at different population sizes, from $10^3$ up to $10^7$. We will compare the results and the average computation time needed using the SS and MS representations.

**Implementation of the SS and MS models**

Both implementations were performed from an object oriented perspective in the C++ programming language. Under the classical SS model each individual stores a DNA sequence of a given, fixed, length. Mutation, recombination and fitness evaluation are modeled as in previous models [1, 35] for a model incorporating recombination). The SS model used here is just an extension of the above two models to incorporate variable population size and the specific progression, i.e. infection, mutation, recombination and surviving of the new generation, of the biological model depicted in the algorithm section.

Under the MS model, we first store the initial sequence as an string called 'Master' and then each individual is defined as an object of the class Subject. Each non-empty object of this class will just store a fitness value and a vector of positions and codons. If the individual has no mutations then the object is empty and has the fitness corresponding to the Master sequence. In any other case, each individual stores a vector of starting codon positions. These positions are stored because the codons they capture have a mutation with respect to the original codon. The corresponding new mutant codon must also be stored for each position in the vector. We will briefly explain how to implement specific evolutionary features using this MS representation.

*Mutation:* For each individual, a number of mutants are randomly generated from a Poisson distribution with parameter $L \times \mu$, where *L* is the sequence length and $\mu$ is the mutation rate per site and generation. These mutants will be distributed by randomly choosing different positions along the sequence. For each new mutant position *mp*, it must be checked if it affects to any previously stored start codon position in the subject. This is easily done just by checking if the value *mp* – *mp* %3, where % stands for the module of integer division, is already present in the vector of positions. If it is not, then just add the mutant position (*mp* – *mp* %3) and the new codon to this subject. Else, if it is already in the vector, this is a recurrent mutation and we must re-compute the new codon. Then, if it reverts to wild, delete this position from the vector; if not, keep the position and substitute the old codon with the new one.

*Recombination*: The MS model provides an intuitive and efficient way to perform recombination. First, given two different proviruses in the same cell, we only must consider mutant positions, i.e. position values that are in any of the two parental individual vectors. As with mutation, we get the number of recombinant points from a Poisson random generator with parameter $L \times r$ and distribute these points randomly along the sequence. Now consider the production of one of the two possible recombinant proviruses. For individual one, we just need to count, for each start codon position in its vector, the number of recombination points before such position. Only if the number is even, including zero, we do include the mutant position in a new offspring vector (see Figure 1). Now we repeat the operation for individual two but only including the positions with an odd number of recombinant points below each position in the vector. We get in this way one of the putative recombinants, the other will be simply the inverse one including odd and even positions, instead of even and odd, for individual one and two, respectively. The only problem with this algorithm is that a recombinant point could fall just within the codon, i.e. given the position $p$, the recombinant point could have a value of $p$ that will break the codon changing from the second position or a value of $p+1$ that will break changing the third codon position. If any of these situations occur we must check if the codon is also a mutant in individual 2 or not, and re-compute the new recombinant codon in consequence. For example, imagine that in individual one, the starting codon position $X$, which is not included in the vector of individual two, is broken changing from the second nucleotide. In this case, we need just to substitute the values $X[2] = $ Master$[X][2]$ and $X[3] = $ Master$[X][3]$ where $X[i]$ refers to the $i$ position of codon $X$ and Master$[X][i]$ stands for the $i$ position of codon $X$ in the Master sequence.

*Fitness evaluation*: For each individual, the initial fitness is set to one. Then, for each selective codon, we check if its position appears in the vector of mutations. If it does not appear then multiply fitness by $1 - s$, else check if the codon is the favored one, if not, then multiply fitness by $1 - s$.

*Model Extensions*: The above model can be easily extended to deal with more complex situations. The extension to diploids is straightforward; we just need to keep two vectors instead of one for each individual. The computation of homozygous or heterozygous states does not pose major problems. We can also consider several individuals instead of one at the initial population. In this case, we must compute a consensus sequence and then representing each original sequence as mutant deviations from the consensus one. We also can consider different populations with different migration rate between them. In this case, we must compute a consensus sequence for each population and recompute each migrant individual as mutant deviations from the consensus of the receiver population. Alternatively, we can compute a metapopulation consensus and then deviate every individual from that.

**Testing**

We study the emergence of HIV drug resistance under therapy conditions. Thus, we evolve a virion population during 250 generations studying the mutant distribution and the increase of average population fitness due to the appearance of known mutations contributing to resistance. We run the simulations in a computer cluster with 13 processors and in a personal computer with Pentium processor. For any case studied it is guaranteed that the processes dispose of 100% of processor time. Concerning the evolution of resistance and the corresponding fitness increase, there were no significant differences between SS and MS models. With population sizes from $N = 10^3$ to $10^5$, we do not observe an increase of fitness or this was very slight. With population size equal to $10^6$, one or two resistance mutations, varying by replicate, were fixed with the corresponding increase of fitness. Full resistance only appears with population size of $10^7$ and only in 20% of the replicates. The population average fitness by replicate in this later case was 0.6 (see continuous line in Figure 2). With this population size we only performed simulations under the MS model due to the high computation cost of this case with the SS model. In the time interval assayed (250 generations) the results did not vary with the different recombination values used. In addition, we obtained the average distribution of the frequency of selective mutations in the population. As expected, with low and medium population sizes, almost all individuals carried no mutations, and a few ones carried just one or two mutations at most. With population sizes higher than $10^5$, i.e. progeny numbers of $10^6$ or higher, the shape of the mutant distribution changes drastically and all viruses carried one or more mutations in the p51 coding region (see bars in Figure 2).

In Figure 3 we can observe the average time by replicate for SS and MS models at the different population sizes. Clearly, the same result is always obtained equal or faster in the MS model than in the SS one, the differences increasing as we increased the population size, implying more than two orders of magnitude with $10^6$ or higher population sizes. The time estimated for the SS model with population size of $10^7$ is a minimum estimate taking into account a linear increase with respect of the $10^6$ case. However, because of memory overflow we were not able to run the SS model with a population size of $10^7$ neither in a personal computer nor in the cluster.

## Discussion

**The ms representation**

We have developed a simple framework that allows the efficient and flexible computer modeling of complex evolutionary situations. The MS methodology used to perform the forward simulations produces the same results than the SS one, but more than two orders of magnitude faster. We have illustrated this with an example that is far from being the optimal situation for the MS model, that is, we have considered a short fragment of DNA, just 3Kb, with a high mutation rate and, even in such situation, the MS model has demonstrated its higher efficiency. Indeed, we have also been able to run the MS model with a population size of $10^7$ both in a cluster and in a personal computer with a Pentium processor in just a few hours. However, optimal modeling situations will be those with high sequence or genome sizes and medium-low mutation rates. In such situations, the MS model should overcome by several orders of magnitude the more classical forward ones. For example, we have been able to evolve in a few seconds on a personal computer ten thousand individuals during 100 generations with diploid genome size of $10^7$ nucleotides with a mutation rate per haploid genome and generation of $U = 0.5$ using the same MS model developed here (not shown). This genome size could be considered a valid approximation to the Drosophila genome discarding the third codon nucleotides. In this context of genome simulation we can predict the substitution rate [36] and when a number of mutations has been fixed, we can compute a second consensus to store the fixed mutations. In this way we should maintain the efficiency of the MS model during a large number of generations. For example, with a gamete mutation rate $U = 0.5$ we expect a substitution rate of 0.5 per diploid genome per generation for neutral or quasi-neutral mutations, which implies about 50 mutants fixed after 100 generations. Therefore, we can decide to update the individual vectors each 100 generations to deal with the redundant information that is being generated due to the evolutionary process.

**The simulation experiment**

Concerning the test case studied, we have used a real sequence and real codon positions and mutations, to see how resistance could evolve in HIV. The mutation space representation is particularly suited to simulate HIV in the framework of the quasispecies theory. This is due to its structure, i.e. a master sequence and a population of individuals represented as deviations from it, and efficiency, allowing for simulating large progeny numbers in the computer. It is in this context where the virus ability to increase its fitness is a direct consequence of the large number of mutant progeny that allows an efficient exploration of the sequence space. The result we have obtained fits perfectly with this quasispecies vision. The larger the population the larger the fitness gain, which coincides with the experimental observation in RNA viruses [37-39].

Of course the model we have used is a simplification, as every model is. We have assumed just one infection rate and one cell type but infection rates could varie depending on the cell type, and there are different cell types with different half-lives that should affect the results of the model. However, double infection rates seem not to be very important here since recombination had no impact onto the fitness increase.

Somewhat different to a previous study [34] we do not detect any recombination effect favoring fixation time. This is due to the time limit imposed of only 250 generations of evolution and the major number of resistant mutants needed. Thus, if we do not impose a time limit or reduce the number of alleles or the mutations necessary for achieving resistance we observe (not shown), as expected, that recombination favors resistance fixation. However, for multiple resistance appearing in short periods of time the key factor seems to be the population size. If this is not large enough, the sequence space cannot be explored sufficiently for the necessary mutations to appear and recombination has nothing to do. Therefore, this is concerned with the current discussion about effective population size ($N_e$) in HIV-1 [40]. Our result seems to corroborate the argument that there is no universally applicable unique value of effective population size for HIV-1 but this will depend on the specific process of interest under study. In the case of the evolution of resistance in a short time period, it seems that no low or medium effective population sizes as $10^3 - 10^5$ should be assumed though in other situations it could [41].

We have taken advantage of the MS representation under a specific forward Montecarlo simulation. The MS could be applied, however, to other kind of forward simulations as, for example, the direct method of Gillespie [42] in the framework of the stochastic time-evolution equation. This method uses Montecarlo simulation to solve numerically the stochastic equation. The MS representation could be used provided that the objects under study could be represented as a deviations from a master or consensus one which do not seems to be the case when simulating chemical reactions [43] but may be possible in the case of viral dynamics [44].

## Conclusions

We propose a forward simulation framework to represent individuals just as the mutations they carry with respect to the wild genotype. The idea is straightforward. However, it has not been used in previous forward simulation models [45-48]. Hence, we demonstrate the efficiency of the MS model by comparing its lower cost in computation space and time with respect to a typical forward implementation. It is concluded that the MS seems to be an efficient way for forward modeling of genomes and populations in the computer.

## Methods

### Model for the simulation experiment

The HIV evolution model used in the simulation experiment to test the MS method is an adaptation of a previous one [32] and modified in [34]. Thus, we further adjust this model to manage real or simulated DNA sequences. Each DNA sequence of length $L$ is a provirus. From the distribution of proviruses we get distributions of two different kinds of cells i.e. single and double infected cells with probabilities $1 - f$ and $f$ respectively. The life cycle takes place as explained in the algorithm section. HIV virions are released from cells and selection acts to allow or not the virions survival. If the cell was double infected, the fitness is evaluated as the average of the two proviruses. At each generation the process of generating new virions from the infecting provirus is repeated $N$ times, so that the population has a variable size with a maximum of $N$ virions.

The fitness of a provirus depends on the selective(s) site(s) defined in the original sequence and on the selective coefficient acting against the virion survival, namely, under therapy conditions the fitness will be 1 only if some specific amino acids are present at given sites of the sequence. Therefore, this extended model includes the calculation of the corresponding amino acids from specific nucleotide positions.

Zidovudine resistance is conferred by the presence of subsets of four or five amino acid substitutions in the HIV-1 reverse transcriptase. However, the dominant resistance phenotype for Zidovudine seems to be Leu 41 and Tyr or Phe 215 [49]. In addition, Val 74 contributes to resistance to Abacavir and Didanosine [50, 51]. To test the two implementations of the model above we will use the coding region of the subunit p51 of the reverse transcriptase from the consensus B_FR.HXB2 HIV-1 sequence to evolve in the computer with mutation and recombination. Fitness will be 1 if and only if a virion has Leu-41, Tyr/Phe-215 and Val-74. On the contrary, fitness will be computed under a multiplicative model multiplying by 1- $s$ for each allele without resistance mutations. Thus, the initial population, without resistance mutations have fitness $(1-s)^3$ which implies a value of 0.125 with the parameter value used of $s = 0.5$.

**Computer simulations**

We used the following parameters in the simulations: Generation time in HIV-1 is considered to be of 1.8 days [52] but see [53]. Therefore, we run 250 generations, which in any case implies more than one year of HIV evolution. With the MS model we run 100 replicates except for population size $10^7$ that run only 10 replicates. With the SS model we run only 10 replicates because of the larger computation time needed. Population size ($N$) was $\{10^3, 10^4, 10^5, 10^6, 10^7\}$; mutation rate per nucleotide site and generation ($\mu$) was $3 \times 10^{-5}$ [54]; recombination rate per site and generation ($r$) was $\{3 \times 10^{-4}, 3 \times 10^{-5}, 3 \times 10^{-6}\}$ [55]; double infection rate ($f$) was 0.8 [56] and selection coefficient ($s$) was 0.5.

# Authors' contributions

AC-R had the original idea for the work, designed and implemented the algorithms and wrote the manuscript.

# Acknowledgements

I wish to thank A. Caballero, E. Rolán-Álvarez, S.T. Rodríguez-Ramilo and H. Quesada and two anonymous reviewers for valuable comments on the manuscript. I am currently funded by an Isidro Parga Pondal research fellowship from Xunta de Galicia (Spain).

# Figures

**Figure 1 - How the algorithm produces one recombinant type after recombination with two break points, one parental provirus having 3 mutants and the other having no mutants (it does not appear in the figure). The number of recombination break points before each mutant position appears between brackets. See text for details.**

**Figure 2 - Selective mutation frequency distribution and fitness gain after 250 generations for different population sizes.**

LogN: Logarithm of population size; Continuous line: represent the fitness gain after 250 generations for each population size. Bars: The frequency of individuals carrying a number of selective mutations (classes from 0 to 3 selective mutations). From left to right, empty bars represent class of 0 selective mutations, bars with vertical lines: 1 selective mutant, hatched lines: 2 selective mutants, filled: 3 selective mutants.

**Figure 3 - Computation time in minutes needed to simulate 250 generations under different population sizes using the SS and MS models.**

LogN: Logarithm of population size; Continuous line: MS model; dashed line: SS model. Estimated: The result for one replicate with the SS model was not yet obtained after 4 days of computation in a 13-processor cluster and finally the process died due to memory overflow.

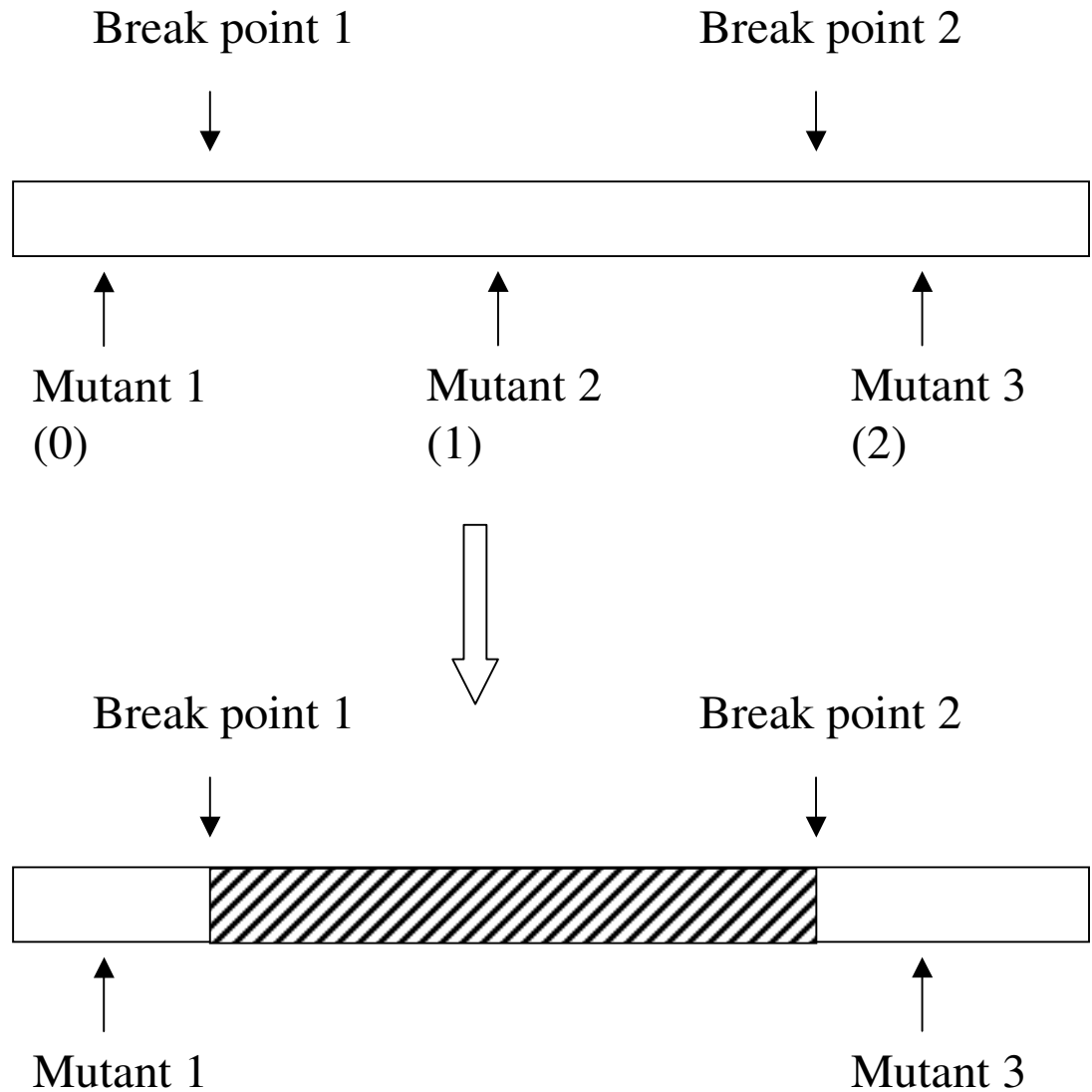

Figure 1

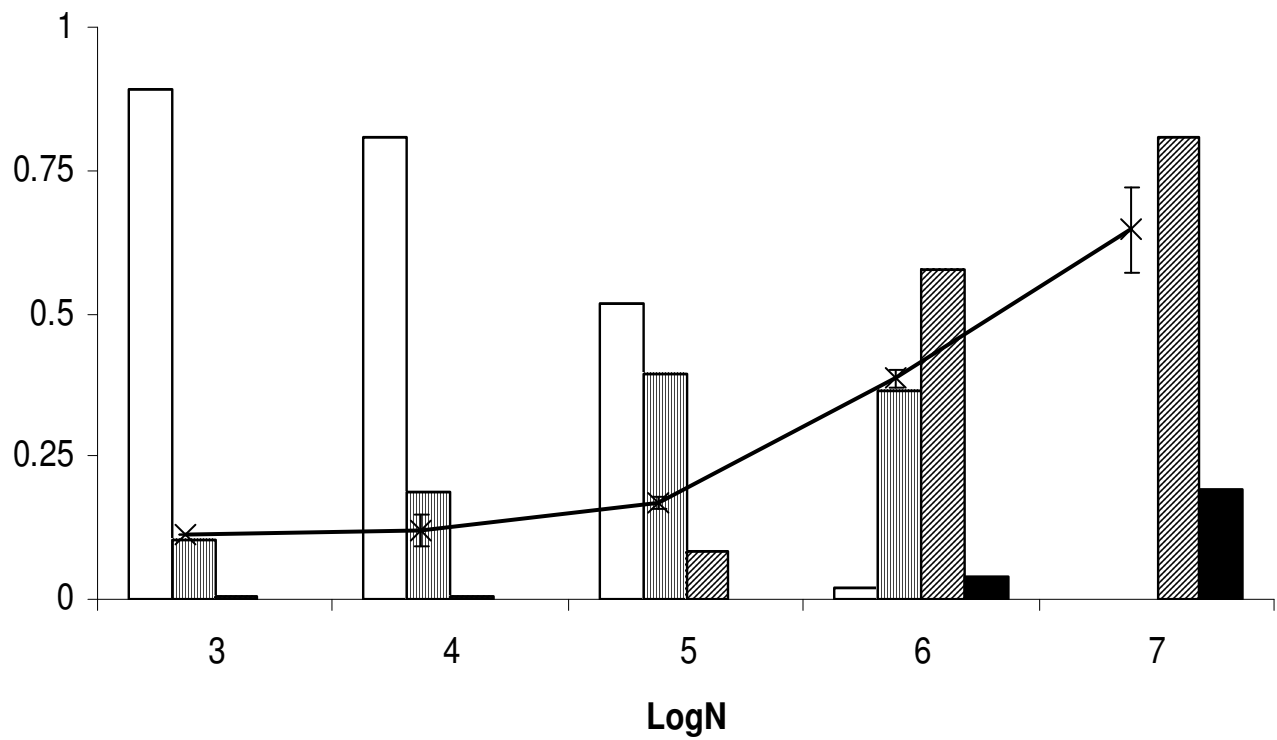

Figure 2

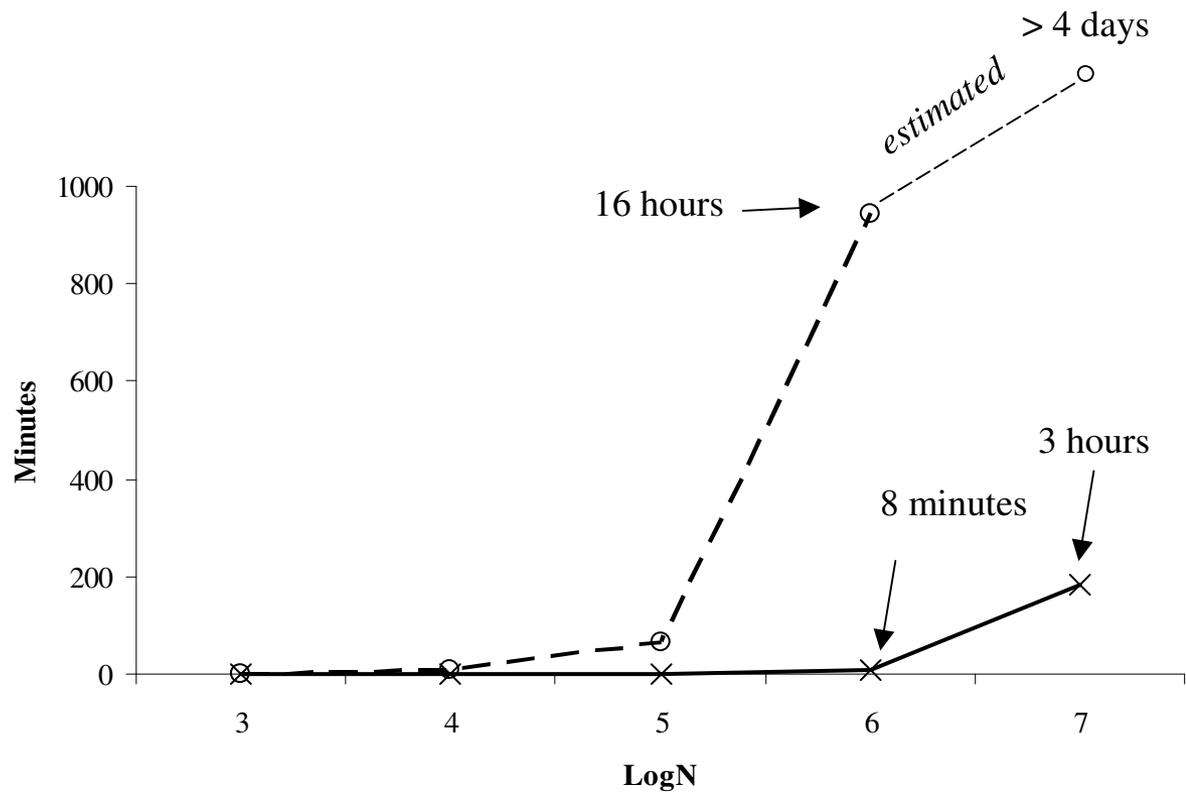

Figure 3